\documentclass[aps,preprint]{revtex4}%
\usepackage{amsfonts}
\usepackage{amsmath}
\usepackage{amssymb}
\usepackage{graphicx}%
\providecommand{\U}[1]{\protect\rule{.1in}{.1in}}
\begin{document}
\title[ ]{Thermal Radiation Equilibrium: (Nonrelativistic) Classical Mechanics versus
(Relativistic) Classical Electrodynamics}
\author{Timothy H. Boyer}
\affiliation{Department of Physics, City College of the City University of New York, New
York, New York 10031}

\begin{abstract}
Physics students continue to be taught the erroneous idea that classical
physics leads inevitably to energy equipartition, and hence to the
Rayleigh-Jeans law for thermal radiation equilibrium. \ Actually, energy
equipartition is appropriate only for \textit{nonrelativistic} classical
mechanics, but has only limited relevance for a \textit{relativistic} theory
such as classical electrodynamics. \ In this article, we discuss
harmonic-oscillator thermal equilibrium from three different perspectives.
\ First, we contrast the thermal equilibrium of nonrelativistic mechanical
oscillators (where point collisions are allowed and frequency is irrelevant)
with the equilibrium of relativistic radiation modes (where frequency is
crucial). \ The Rayleigh-Jeans law appears from applying a dipole-radiation
approximation to impose the nonrelativistic mechanical equilibrium on the
radiation spectrum. \ In this discussion, we note the possibility of
zero-point energy for relativistic radiation, which possibility does not arise
for nonrelativistic classical-mechanical systems. \ Second, we turn to a
simple electromagnetic model of a harmonic oscillator and show that the
oscillator is fully in radiation equilibrium (which involves \textit{all}
radiation multipoles, dipole, quadrupole, etc.) with classical electromagnetic
zero-point radiation, but is \textit{not} in equilibrium with the
Rayleigh-Jeans spectrum. \ Finally, we discuss the contrast between the
flexibility of nonrelativistic mechanics with its arbitrary potential
functions allowing separate scalings for length, time, and energy, with the
sharply-controlled behavior of relativistic classical electrodynamics with its
single scaling connecting together the scales for length, time, and energy.
\ It is emphasized that within classical physics, energy-sharing,
velocity-dependent damping is associated with the low-frequency,
nonrelativistic part of the Planck thermal radiation spectrum, whereas
acceleration-dependent radiation damping is associated with the high-frequency
adiabatically-invariant and Lorentz-invariant part of the spectrum
corresponding to zero-point radiation. \ 

\end{abstract}
\maketitle

\section{Introduction}

\subsection{Erroneous Textbook Claim}

The current textbooks of modern physics still make the erroneous claim that
classical physics leads inevitably to the Rayleigh-Jeans spectrum for thermal
radiation equilibrium.\cite{modern} \ Indeed the treatment of classical
physics in these texts seems to have progressed only so far as James Jeans'
\textit{Report on Radiation and the Quantum-Theory} in 1914, but no
further.\cite{Jeans} In this article, we make another\cite{BB2018} attempt to
counteract the erroneous textbook claim. \ \ Specifically, we point out that a
simple electromagnetic-model harmonic oscillator is \textit{not} in
equilibrium with the Rayleigh-Jeans spectrum. \ In the discussion, we
emphasize the contrasting points of view between nonrelativistic classical
mechanics and relativistic classical electrodynamics in connection with
equilibrium for harmonic oscillators. \ 

Our earlier attempt\cite{BB2018} to correct the misinformation regarding
thermal radiation equilibrium in classical physics involved an extended
historical survey. \ It was pointed out that the introduction of classical
electromagnetic zero-point radiation led to modifications of the historical
arguments. \ The modified arguments provide natural classical explanations for
the Planck spectrum within relativistic classical physics. \ The Planck
spectrum with zero-point radiation corresponds to a radiation energy per
normal mode
\begin{equation}
U_{rad}(\omega,T)=\frac{1}{2}\hbar\omega\coth\left(  \frac{\hbar\omega}%
{2k_{B}T}\right)  =\frac{\hbar\omega}{\exp\left[  \hbar\omega/\left(
k_{B}T\right)  \right]  -1}+\frac{1}{2}\hbar\omega. \label{Pla}%
\end{equation}
The spectrum involves a transition from the relativistic high-frequency
region, $\hbar\omega/\left(  k_{B}T\right)  >>1$, associated with
adiabatically-invariant and Lorentz-invariant zero-point radiation where
$U_{rad}\left(  \omega,T\right)  \rightarrow\left(  1/2\right)  \hbar\omega$,
over to the low-frequency Rayleigh-Jeans region, $\hbar\omega/\left(
k_{B}T\right)  <<1$, associated with nonrelativistic energy-sharing behavior
where $U_{rad}\left(  \omega,T\right)  \rightarrow k_{B}T$. \ 

\subsection{Three Analyses for Radiation Equilibrium}

The present discussion is quite different from the previous historical survey.
\ Here we present three different analyses of radiation equilibrium connected
to harmonic oscillator systems; these analyses include aspects of
point-particle collisions, adiabatic invariance, scattering, and scaling which
do not appear in earlier work. \ 

\subsubsection{Point-Particle Collisions and Adiabatic Invariance}

After some basic preliminaries, we start our analysis by contrasting the
equilibrium for two different situations; one is a box containing a finite
number of mechanical oscillators connected through point collisions with a
free particle, and the other involves a charged mechanical oscillator coupled
to a divergent number of electromagnetic radiation modes. \ When the
equipartition ideas of nonrelativistic mechanics are imposed upon radiation
treated in the dipole approximation, the Rayleigh-Jeans spectrum appears.
\ Our analysis also suggests the possibility of classical zero-point energy
for radiation associated with oscillator adiabatic invariance , something
which is not possible for classical mechanical systems which allow
point-particle collisions.

\subsubsection{Equilibrium Under Scattering}

In the second equilibrium analysis, we emphasize that a charged harmonic
oscillator connected to radiation through the dipole approximation can come to
equilibrium with \textit{any} radiation spectrum. \ In order to obtain a
specific radiation spectrum, some additional assumptions are needed. \ The
Rayleigh-Jeans spectrum and zero-point spectrum correspond to the extremes
allowed by Wien's displacement theorem. \ If we turn to invariance under
scattering as our criterion for a preferred spectrum, then\ again there are
two natural possibilities. \ In one possibility, the mechanical behavior is
modified to an \textit{anharmonic} nonrelativistic oscillator while
maintaining the \textit{dipole} approximation to radiation. \ The other
possibility involves maintaining the \textit{harmonic} mechanical behavior but
going to \textit{electromagnetic interactions} \textit{beyond the dipole}
approximation, giving an approximately relativistic oscillator system. \ We
point out that the first nonrelativistic procedure leads to the Rayleigh-Jeans
spectrum, while the second relativistic analysis leads to the classical
zero-point radiation spectrum. \ For a harmonic oscillator, the Rayleigh-Jeans
spectrum arising from the energy equipartition ideas of nonrelativistic
classical mechanics is not tolerated beyond the dipole coupling to radiation. \ 

\subsubsection{Scaling in Nonrelativistic Mechanics and in Relativistic
Electrodynamics}

Finally, in the third equilibrium analysis, we point out the sharp differences
in scaling behavior between nonrelativistic mechanics and relativistic
electrodynamics, and connect these differences to the requirements for thermal
equilibrium. The enormous flexibility of nonrelativistic classical mechanics
contrasts with the strictly-controlled radiation connections of relativistic
classical electrodynamics. \ And this contrast is crucial in questions of
thermal radiation equilibrium. \ We emphasize that, within classical physics,
the full Planck spectrum requires both velocity-dependent damping and
acceleration-dependent radiation damping. \ 

\subsection{Missing Aspects in the Physics of 1900: Zero-Point Radiation and
Relativity}

The foundation for the erroneous claims regarding classical thermal radiation
which pervade the textbooks and the internet lies in the fact that the
physicists of the early 20th century \textit{were} (just as the physicists of
today \textit{are}) unaware of two crucial aspects of classical physics
existing within classical electrodynamics: 1)the existence of classical
electromagnetic zero point radiation, and 2)the importance of special
relativity. \ In this article, we illustrate the importance of these two
aspects of classical physics by comparing the ideas of thermal equilibrium
which arise in classical mechanics and in classical electrodynamics. \ The
models for our discussion are harmonic oscillators.

\section{Basic Preliminaries}

\subsection{Some Contrasts Between (Nonrelativistic) Classical Mechanics and (
Relativistic) Classical Electrodynamics.}

In order to prepare the reader for our analysis, we first mention some of the
important contrasts between classical mechanics and classical electrodynamics.
\ Classical mechanics allows point collisions between massive particles which
provide velocity-dependent damping, whereas classical electrodynamics involves
only long-range forces associated with electromagnetic fields. \ Classical
mechanics allows particles to exist without any connection to a
spacetime-dependent field, whereas in classical electrodynamics, every charged
particle is connected to electromagnetic fields which introduce
acceleration-dependent radiation forces. \ There is no special role for
oscillation \textit{frequency} when considering thermal equilibrium within
nonrelativistic classical mechanics, whereas classical electrodynamics
determines thermal equilibrium through the frequency-dependent connection of
charges to radiation fields. \ Finally, the \textit{relativistic} theory of
classical mechanical particles is an insignificant part of mechanics; the
\textit{relativistic} theory is restricted to point collisions between
particles and allows no potential functions.\cite{CJS} Classical mechanics
which allows the interaction of particles beyond point collisions satisfies
Galilean symmetry, whereas classical electrodynamics satisfies Lorentz symmetry.

\subsection{Harmonic Oscillator in Action-Angle Variables}

A one-dimensional harmonic oscillator of natural (angular) frequency
$\omega_{0}$ and mass $m$ can be described by the Hamiltonian
\begin{equation}
H\left(  x,p\right)  =p^{2}/\left(  2m\right)  +\left(  1/2\right)
m\omega_{0}^{2}x^{2}=U_{osc}. \label{Ham}%
\end{equation}
Within nonrelativistic mechanics, this Hamiltonian will give the equations of
motion for the position $x$ and momentum $p$ while the initial conditions will
determine the actual motion, including the energy $U_{osc}$ and phase of the
oscillation. \ When discussing adiabatic changes in the oscillator's natural
frequency $\omega_{0}$, it is convenient to reexpress the Hamiltonian in terms
of action-angle variables $w$ and $J$ using the canonical transformation with
the connection%

\begin{equation}
x=\sqrt{\frac{2J}{m\omega_{0}}}\sin w\text{ \ and \ \ }p=\sqrt{2m\omega_{0}%
J}\cos w.
\end{equation}
In terms of the new variables $w$ and $J$, the Hamiltonian takes the form
\begin{equation}
H=\frac{p^{2}}{2m}+\frac{1}{2}m\omega_{0}^{2}x^{2}=J\omega_{0}=U_{osc}.
\label{HamJ}%
\end{equation}
In terms of action-angle variables, the oscillator Hamiltonian no longer
refers to the particle mass $m$ nor the amplitude of oscillation $x$. \ The
initial conditions again give the oscillator energy $U_{osc}=J\omega_{0}$ and
the initial phase $\phi$ in $w=\omega_{0}t+\phi$.

\subsection{Thermodynamics of the Harmonic Oscillator and Wien's Theorem}

A harmonic oscillator (whether a mechanical oscillator or a radiation mode) is
such a simple system that its thermodynamic behavior depends upon only two
variables, its natural frequency $\omega_{0}$ and its temperature $T$. \ By
considering the adiabatic change of the oscillator frequency $\omega_{0}$, it
is easy to derive Wien's displacement theorem indicating that in thermal
equilibrium, the energy of the oscillator must take the form\cite{Wein}%
\begin{equation}
U_{osc}(\omega_{0},T)=\omega_{0}f\left(  \omega_{0}/T\right)  , \label{WienD}%
\end{equation}
where $f\left(  \omega_{0}/T\right)  $ is an unknown function of the ratio
$\omega_{0}/T$. \ The extremes of energy equipartition and zero-point energy
correspond to the limiting situations where the energy $U_{osc}(\omega_{0},T)$
becomes independent of one of the two variables. \ Thus if $f\left(
\omega_{0}/T\right)  \rightarrow const\times\left(  T/\omega_{0}\right)  $,
then we have energy equipartition $U_{osc}(\omega_{0},T)\rightarrow
const\times T$; on the other hand, if $f\left(  \omega_{0}/T\right)
\rightarrow const^{\prime}$, then one finds temperature-independent zero-point
energy $U_{osci}\left(  \omega_{0},T\right)  \rightarrow const^{\prime}%
\times\omega_{0}$. \ If we consider only the first two laws of thermodynamics,
then the function $f\left(  \omega_{0}/T\right)  $ giving the transition
between these extremes is unknown. \ Only if we add further criteria such as
"smoothness"\cite{Wein} or inclusion of the third law,\cite{third} do we
obtain the Planck spectrum with zero-point energy given in Eq. (\ref{Pla}).
\ However, both these additional criteria are not currently in fashion in
physics, and below we will consider other possibilities.

\subsection{Thermal Radiation Modes}

The solutions of Maxwell's equations involve electromagnetic fields arising
both from charge and current sources, and also from boundary conditions. \ The
boundary-condition electromagnetic fields satisfy the source-free Maxwell
equations. \ Thermal radiation can be treated as boundary-condition radiation,
and can be expanded in terms of plane waves with random phases satisfying
periodic boundary conditions in a large cubic box with sides of length $a,$%
\begin{equation}
\mathbf{E}(\mathbf{r},t)=%
%TCIMACRO{\dsum _{\mathbf{k,\lambda}}}%
%BeginExpansion
{\displaystyle\sum_{\mathbf{k,\lambda}}}
%EndExpansion
\widehat{\epsilon}(\mathbf{k},\lambda)\left(  \frac{8\pi U_{rad}(\omega
)}{a^{3}}\right)  ^{1/2}\frac{1}{2}\left\{  \exp\left[  i\mathbf{k}%
\cdot\mathbf{r}-i\omega t+i\theta\left(  \mathbf{k},\lambda\right)  \right]
+cc\right\}  , \label{Eran}%
\end{equation}%
\begin{equation}
\mathbf{B}(\mathbf{r},t)=%
%TCIMACRO{\dsum _{\mathbf{k,\lambda}}}%
%BeginExpansion
{\displaystyle\sum_{\mathbf{k,\lambda}}}
%EndExpansion
\widehat{\mathbf{k}}\times\widehat{\epsilon}(\mathbf{k},\lambda)\left(
\frac{8\pi U_{rad}(\omega)}{a^{3}}\right)  ^{1/2}\frac{1}{2}\left\{
\exp\left[  i\mathbf{k}\cdot\mathbf{r}-i\omega t+i\theta\left(  \mathbf{k}%
,\lambda\right)  \right]  +cc\right\}  , \label{Bran}%
\end{equation}
where the sum over the wave vectors $\mathbf{k}=\widehat{x}2\pi
l/a+\widehat{y}2\pi m/a+\widehat{z}2\pi n/a$ involves integers $l,m,n=0,\pm
1,\pm2,...$ running over all positive and negative values, there are two
polarizations $\widehat{\epsilon}(\mathbf{k},\lambda),$ $\lambda=1,2,$ and the
random phases $\theta(\mathbf{k,\lambda})$ are distributed uniformly over the
interval $(0,2\pi],$ independently for each wave vector $\mathbf{k}$ and
polarization $\lambda$. \ The notation \textquotedblleft$cc$\textquotedblright%
\ refers to the complex conjugate. \ We have assumed that the radiation
spectrum is isotropic, and that the energy per normal mode at radiation
frequency $\omega=ck$ is given by $U_{rad}(\omega)$. \ There are a divergent
number of radiation normal modes with no limit on the frequency $\omega.$ \ 

\subsection{Equilibrium Between a Charged Oscillator and Random Radiation in
the Dipole Approximation}

When treated in the dipole approximation, the interaction of a charged
harmonic oscillator with the random radiation in Eqs. (\ref{Eran}) and
(\ref{Bran}) can be written as Newton's second law%
\begin{equation}
m\ddot{x}=-m\omega_{0}^{2}x+m\tau\dddot{x}+eE_{x}\left(  0,t\right)  ,
\end{equation}
involving the harmonic restoring force $-m\omega_{0}^{2}x$, the dipole
radiation damping force $m\tau\dddot{x}$ with $\tau=2e^{2}/\left(
3mc^{3}\right)  $, and the radiation driving force treated in dipole
approximation $eE_{x}\left(  0,t\right)  $. \ This equation of motion has been
solved many times, beginning with Planck's work at the end of the 19th
century.\cite{Planck}\cite{M63}\cite{Rev75}\cite{Lav} \ It is found that in
this dipole approximation, the oscillator acts essentially like a radiation
mode at the oscillation frequency $\omega_{0}$: the average energy of the
oscillator matches the average energy of the radiation modes at frequency
$\omega_{0}$, $U_{osc}\left(  \omega_{0}\right)  =U_{rad}\left(  \omega
_{0}\right)  $, and the phase space distribution $P_{osc}(x,p)$ of the
oscillator takes the form%
\begin{equation}
P_{osc}\left(  x,p\right)  =const\times\exp\left[  -H\left(  x,p\right)
/U_{rad}\left(  \omega_{0}\right)  \right]  \label{PHU}%
\end{equation}
where $H\left(  x,p\right)  $ is the Hamiltonian of the oscillator given in
Eqs. (\ref{Ham}) or (\ref{HamJ}). \ 

In the dipole approximation, the harmonic oscillator is in equilibrium with
\textit{any} spectrum $U_{rad}\left(  \omega\right)  $ of isotropic random
radiation. \ Thus, other than the requirement of isotropy in direction, the
spectrum $U_{rad}\left(  \omega\right)  $ is arbitrary. \ The oscillator will
enforce isotropic behavior for the radiation spectrum, but the harmonic
oscillator will not change the frequency distribution of radiation among the
various frequencies.\cite{Rev75}

\section{Exploring Equilibrium for Oscillators in Mechanics and in
Electromagnetic Radiation}

We now turn from the preliminary aspects of our analysis over to specific
treatments of radiation equilibrium for harmonic oscillators. \ 

\subsection{Equilibrium for Systems of Oscillators}

Suppose that we have an uncharged free particle of mass $M$ and a special
harmonic oscillator of (angular) frequency $\omega_{0}$ at our disposal. \ We
wish to contrast the equilibrium behavior for two different systems. \ The
first system involves a collection of $N$ non-interacting,
classical-mechanical, one-dimensional oscillators of various frequencies
$\omega$ in an elastic-walled box; the second involves the electromagnetic
radiation in a reflecting-walled box. \ Both boxes can be described as
containing harmonic oscillators inside, the one in terms of mechanical
oscillators and the other in terms of electromagnetic radiation modes. \ We
assume that the oscillators in both boxes are in \textquotedblleft thermal
equilibrium,\textquotedblright\ and we wish to explore this thermal
equilibrium. \ 

\subsection{Equilibrium for Mechanical Oscillators}

\subsubsection{Equilibrium Through Point-Particle Collisions}

If we introduce our special oscillator into the box of $N$ mechanical
oscillators, the oscillator will not come to equilibrium in the box unless
there is some interaction between this oscillator and the original oscillators
in the box. \ However, if we now introduce our free particle $M$ into the box
of mechanical oscillators, then the particle will collide with all the masses
of the mechanical oscillators and will bring to equilibrium the entire
collection (including both our special oscillator, the $N$ oscillators in the
box, and also the mass $M$) by sharing the total system energy $U_{total}%
$\ among the finite number $N+1$\ of mechanical oscillators and the mass $M$.
\ Each oscillator of mechanical frequency $\omega$ will come to equilibrium at
the Boltzmann distribution $P_{osc}\left(  J,\omega\right)  =const\times
\exp\left[  -J\omega/U_{av}\right]  $ where here $J$ is the action variable of
the oscillator with Hamiltonian $H=J\omega$, and $U_{av}=U_{total}/(N+1+3/2)$
is the average energy, (from energy equipartition) with the average energy of
each one-dimensional oscillator $2/3$ that of the mass $M$ which has
three-dimensional motion. \ We notice that the natural frequency $\omega$ of
an oscillator is irrelevant for the average oscillator energy at equilibrium;
our special oscillator at frequency $\omega_{0}$ has the same average energy
as any oscillator at any frequency $\omega.$ However, we notice that the phase
space distribution $P_{osc}\left(  J,\omega\right)  $ for an oscillator does
depend upon its frequency $\omega$ and also upon the average energy $U_{av}$
for the entire system.

\subsubsection{Change of Equilibrium Due to Adiabatic Change of Oscillator
Frequency}

If we now isolate our special oscillator and carry out an adiabatic change in
the frequency, the quantity of energy divided by frequency, $U_{av}/\omega
_{0}=J$, is an adiabatic invariant and does not change. \ When the frequency
is increased from $\omega_{0}$ to $\omega_{0}^{\prime}$, the energy of our
special oscillator is increased from $U_{av}$ over to $U^{\prime}=\left(
\omega_{0}^{\prime}/\omega_{0}\right)  U_{av}$. \ This energy $U^{\prime}$ is
now above the average energy $U_{av}$ of the other oscillators in the box. On
reconnecting our special oscillator to the collection of the other mechanical
oscillators and the mass $M$, the average energy of each oscillator will
increase, and the phase space distribution of each oscillator will change to
give a larger average value of its action variable $J$. \ Thus an adiabatic
change in the frequency of our special oscillator will indeed disturb the
equilibrium of the entire system.

\subsection{Equilibrium for Random Radiation}

\subsubsection{Charged-Oscillator Equilibrium Depends on Oscillator Frequency}

This thermal-equilibrium situation for nonrelativistic mechanical oscillators
(where point-particle collisions are allowed, and there is no particular role
regarding energy equilibrium for the oscillator frequency) is to be contrasted
with the electromagnetic situation which involves a divergent number of
massless time-harmonic radiation modes characterized by frequency. \ In
analogy with the mechanical case, the introduction of a special harmonic
oscillator of frequency $\omega_{0}$ into a conducting-walled enclosure gives
us no information unless we specify the interaction of the oscillator with the
radiation modes of the enclosure. \ A free but uncharged particle $M$ has no
interaction with the electromagnetic oscillators. \ However, we can connect
our special mechanical oscillator to radiation by taking the oscillating
particle as charged. \ As pointed out by Planck at the end of the 19th century
and noted above in our preliminaries, treated in the dipole approximation,
this electromagnetic harmonic oscillator will interact with the radiation
modes at the same frequency as the natural frequency of the oscillator and
will come to equilibrium with these modes, but not with modes of differing
frequency. \ Thus the frequency-dependent connection between an oscillator and
electromagnetic radiation is a crucial aspect of electromagnetism. \ 

\subsubsection{Possibility of Adiabatic Invariance in Zero-Point Radiation}

We now consider an adiabatic change in the frequency $\omega_{0}$ of our
oscillator over to a new frequency $\omega_{0}^{\prime}$. \ During this
adiabatic change, the ratio of energy to frequency, $U_{0}/\omega_{0}=J,$ of
our special oscillator is unchanged. \ We can now have our oscillator (treated
in the dipole approximation) interact with radiation at the new frequency
$\omega_{0}^{\prime}$. \ If the radiation modes at $\omega_{0}^{\prime}$ have
greater energy than our oscillator, then, on interaction, there will be energy
transferred from the radiation modes at $\omega_{0}^{\prime}$ over to our
special oscillator. \ On the other hand, if the radiation modes at $\omega
_{0}^{\prime}$ have less energy, then the transfer is reversed. \ However, we
also have the possibility that our special oscillator has the same average
energy as the radiation modes at $\omega_{0}^{\prime}$ and so is already in
equilibrium with these modes. \ This situation of undisturbed equilibrium
despite an adiabatic change of an oscillator's frequency and energy is an
entirely new possibility which does not arise for mechanical oscillators where
energy equipartition is involved. \ Here in the electromagnetic situation, we
have the possibility of making adiabatic changes in the frequency and energy
of our oscillator and yet not transferring any average energy between the
electromagnetic radiation modes. \ If this situation of thermal equilibrium
holds for all radiation modes, then the classical electromagnetic spectrum is
Lorentz invariant and is termed \textquotedblleft classical electromagnetic
zero-point radiation.\textquotedblright\cite{ZPE}

\subsubsection{Introduction of Planck's Constant $\hbar$}

The scale of this zero-point radiation involves the ratio $U/\omega$ which is
the same for all radiation frequencies. \ The scale which fits with nature
corresponds to the introduction of Planck's constant $\hbar$ as the scale
invariant so that the average energy $U_{rad}\left(  \omega\right)  $ of a
radiation mode of frequency $\omega$ satisfies $U_{rad}\left(  \omega\right)
=\left(  1/2\right)  \hbar\omega$. \ There is no question that, when
interpreted in terms of classical physics, nature contains classical
electromagnetic zero-point radiation. \ Thus the introduction of Planck's
constant $\hbar$ as the scale of classical zero-point radiation gives fully
classical explanations for Casimir forces, van der Waals forces, oscillator
specific heats, diamagnetism, and the Planck spectrum of thermal
radiation.\cite{Rev} \ Zero-point radiation involves the same phase space
distribution for every radiation mode no matter what its frequency,
\begin{equation}
P_{rad\text{ zp}}\left(  J\right)  =const\times\exp\left[  -\frac{J\omega
}{\left(  1/2\right)  \hbar\omega}\right]  =const\times\exp\left[  -\frac
{J}{\hbar/2}\right]  . \label{Pradzp}%
\end{equation}
\ 

\subsubsection{Planck's Constant $\hbar$ Can Appear in Classical or Quantum
Theories}

The appearance of Planck's constant $\hbar$ within classical physics provokes
the indignation of some physicists. \ Some uninformed physicists continue to
insist that Planck's constant $\hbar$ is a \textquotedblleft quantum
constant,\textquotedblright\ and that any theory in which $\hbar$ appears has
a \textquotedblleft quantum\textquotedblright\ element. \ That this claim is
untrue has been pointed out in work published in the American Journal of
Physics.\cite{const} \ Actually, Planck's constant $\hbar$ is a physical
constant, like Cavendish's constant $G$, and can appear in any theory which
involves an appropriate scale, just as the constant $G$ can appear in both
Newtonian gravity and in general relativity since $G$ sets the scale of the
gravitational interaction. \ In modern theory, Planck's constant appears both
as the scale of action in quantum theory and as the scale of classical
zero-point radiation in classical electrodynamics.

Because Planck's constant $\hbar$ does not appear in Maxwell's equations but
only in the source-free boundary condition\cite{const} on the equations, there
are two natural versions of classical electrodynamics, one including $\hbar$
and one not. \ The version which includes Planck's constant provides a natural
classical explanation of the Planck spectrum\cite{BB2018} and of some other
phenomena\cite{Rev} with a scale set by $\hbar.$ \ The version which appears
in current textbooks of classical electromagnetism assumes that the
source-free electromagnetic field vanishes and so offers no explanation of
phenomena at the scale of $\hbar$. \ 

It remains true that Planck's constant $\hbar$ has no place in classical
\textit{mechanics} because classical mechanics (with point-particle collisions
which share all energy) allows no zero-point energy. \ Of course, classical
mechanics also does not allow massless waves which carry energy and momentum.
\ However, classical electromagnetism is quite different from classical
mechanics. \ Classical electromagnetism includes massless waves, and, indeed,
has a natural place for Planck's constant as the scale of classical
electromagnetic zero-point radiation. \ 

\section{Radiation Equilibrium for an Oscillating Charged System}

\subsection{A Point Harmonic Oscillator Does Not Determine Radiation
Equilibrium}

In the discussion above, we have seen the reappearance of the same two
oscillator energies (equipartition and zero-point energy) which appeared from
the limits of Wien's law (\ref{WienD}), arising from the thermodynamics of the
harmonic oscillator. Again it must be emphasized that since a charged harmonic
oscillator treated in the dipole approximation comes to equilibrium with
\textit{any} radiation spectrum, the determination of the actual spectrum of
thermal radiation equilibrium must be based upon other factors. \ Here we list
several criteria.

\subsection{Specific Criteria for Radiation Equilibrium}

\subsubsection{Adoption of the Nonrelativistic Oscillator Equilibrium}

In the early years of the 20th century, just as in the modern physics texts of
today, it was assumed\cite{modern}\cite{Jeans} that nonrelativistic
statistical mechanics correctly determines the phase space of the harmonic
oscillator according to the Boltzmann distribution%
\begin{equation}
P_{osc}\left(  x,p\right)  =const\times\exp\left[  -H\left(  x,p\right)
/\left(  k_{B}T\right)  \right]  \label{PJB}%
\end{equation}
Then comparing equations (\ref{PHU}) and (\ref{PJB}), it was concluded that
the spectrum of random radiation in equilibrium with the oscillator must be
$U_{rad}\left(  \omega\right)  =k_{B}T$. \ In other words, nonrelativistic
classical mechanics, which uses Boltzmann statistical mechanics, must lead to
the Rayleigh-Jeans spectrum for relativistic classical radiation. \ 

In 1910, Einstein and Hopf\cite{EH}\cite{Milonni} tried to avoid full
Boltzmann statistical mechanics and to use instead only the well-established
average thermal \textit{kinetic} energy of a massive nonrelativistic particle
in one dimension as $\left(  1/2\right)  Mv^{2}=\left(  1/2\right)  k_{B}T$.
\ However, they again arrived at the Rayleigh-Jeans spectrum for thermal
radiation. \ 

\subsubsection{\textit{Adoption of the Spectrum Invariant Under Adiabatic
Change }}

An entirely different criterion involves adiabatic invariance. \ We saw above
that, under an adiabatic change of frequency, a charged harmonic oscillator of
frequency $\omega_{0}$ and energy $U\left(  \omega_{0}\right)  $ (when treated
in the dipole approximation) will not transfer energy among radiation modes if
the radiation spectrum is Lorentz invariant. \ Thus the assumption of
adiabatic invariance for the oscillator picks out a special radiation
spectrum, that of classical electromagnetic zero-point radiation. \ This is
also the spectrum allowed by the limit of Wien's law (\ref{WienD}) which makes
the equilibrium spectrum independent of the temperature $T$. \ 

\subsubsection{\textit{Adoption of the Spectrum Invariant Under Scattering}}

A third criterion involves invariance under scattering. \ The charged harmonic
oscillator treated in the dipole approximation scatters radiation so as to
make the radiation pattern isotropic; however, the scattering by the
oscillator does not change the frequency spectrum of the random radiation.
\ In order to obtain \ a preferred spectrum based upon invariance under
\textit{scattering}, we must go beyond use of the harmonic oscillator treated
in the dipole approximation. \ 

\paragraph{Two Oscillator Extensions Leading to Equilibrium Radiation Spectra}

There are two natural extensions of the harmonic oscillator model. \ The first
possible extension retains the electromagnetic aspect (continuing the use of
the dipole approximation),\ but changes the mechanics (going beyond the
harmonic oscillator over to an anharmonic oscillator). \ The second possible
extension retains the mechanics (continuing the use of a purely harmonic
mechanical oscillator), but changes the electromagnetic connection (going
beyond the dipole radiation approximation over to full electromagnetic
interaction at all harmonics). \ The two possible extensions give different
preferred equilibrium spectra. \ The nonrelativistic mechanical extension
while retaining the dipole radiation approximation gives the Rayleigh-Jeans
spectrum. \ The relativistic electromagnetic extension while retaining the
harmonic mechanical behavior gives classical zero-point radiation. \ Both
extensions will be discussed below. \ 

\subsection{Radiation Scattering by an Anharmonic Nonrelativistic Mechanical
System in the Dipole Approximation}

In order to avoid the use of the ideas of classical statistical mechanics,
various physicists have considered the transition beyond harmonic oscillators
to nonrelativistic anharmonic oscillators which will scatter radiation and
change the spectrum of the random radiation. \ In 1924, Van Vleck\cite{VV}%
\ considered (and partially published) a general analysis of nonlinear
nonrelativistic mechanical systems in scattering equilibrium where the
radiation was treated in the dipole approximation. \ Work by other authors has
considered scattering by a nonrelativistic nonlinear oscillator,\cite{nonlin}
and scattering by nonrelativistic potentials where the charged particle
momentum took relativistic form.\cite{Blanco} \ All of these classical
scattering calculations arrived at the Rayleigh-Jeans spectrum. \ None of
these calculations treats a relativistic scattering system. \ 

A nonrelativistic anharmonic oscillator with a nonlinear term in energy
$\alpha x^{3}$ and Hamiltonian
\begin{equation}
H\left(  x,p\right)  =p^{2}/\left(  2m\right)  +\left(  1/2\right)
m\omega_{0}^{2}x^{2}+\alpha x^{3} \label{Hnonlin}%
\end{equation}
provides a simple example of a scattering system which will enforce the
Rayleigh-Jeans spectrum for radiation equilibrium. \ The equation of motion
takes the form%
\begin{equation}
m\ddot{x}=-m\omega_{0}^{2}x+3\alpha x^{2}+m\tau\dddot{x}+eE_{x}\left(
0,t\right)  ,
\end{equation}
and continues the use of the dipole approximation in the radiation damping
$m\tau\dddot{x}$ and in the electromagnetic force $eE_{x}\left(  0,t\right)
$. \ The mechanical motion of the oscillator now involves a mechanical
oscillation at frequency $\omega_{1}$ which is different from $\omega_{0}$ and
which depends upon the constant $\alpha$ giving the scale of the nonlinear
term.\cite{Born} \ The mechanical motion now includes the harmonics\cite{Born}
of the frequency $\omega_{1}$
\begin{equation}
x\left(  t\right)  =D_{0}+D_{1}\cos\left[  \omega_{1}t+\phi_{1}\right]
+D_{2}\cos\left[  2\omega_{1}+\phi_{2}\right]  +... \label{xtVV}%
\end{equation}
Since the mechanical motion has oscillations at all the harmonics $n\omega
_{1}$ of the fundamental mechanical frequency $\omega_{1}$, the nonlinear
oscillator will interact with radiation in the dipole approximation at all the
harmonics of the oscillation frequency $\omega_{1}$. \ This interaction with
radiation will transfer energy among the radiation modes, and so will lead to
a unique equilibrium spectrum. \ We note that the radiation equilibrium is
determined by the \textit{nonrelativistic mechanical system} and not by any
properties of the radiation with which the system has minimal (dipole)
interaction. \ In all treatments with nonrelativistic nonlinear mechanical
scatterers connected to radiation through the dipole approximation, one finds
the Rayleigh-Jeans spectrum as the unique spectrum of radiation equilibrium.
\ When van Vleck\cite{VV} partially published his analysis in 1924, the work
was regarded as confirming that the Rayleigh-Jeans spectrum was the
appropriate equilibrium radiation spectrum expected within classical physics. \ 

\subsection{ \ Radiation Scattering by a Charged Harmonic Oscillator with Full
Electromagnetic Interactions}

\subsubsection{One-Dimensional Electromagnetic Harmonic Oscillator}

In order to emphasize electromagnetic interactions in contrast to arbitrary
mechanical potentials, we mention here the possibility of a one-dimensional
oscillator where the oscillating particle of mass $m$ and charge $e$ moves
under purely electromagnetic forces. \ We imagine a charge $e$ which is
constrained to move along the straight line between two charges $q$ (of the
same sign as $e$) separated by a distance $l$. \ The charge $e$ will then
oscillate in one dimension along the line between the charges $q,$ but is in
unstable equilibrium against motion to the side if the constraint were
removed. For small oscillations, the mechanical motion of $e$ is harmonic with
(angular) frequency $\omega_{0}=\left[  32eq/\left(  ml^{3}\right)  \right]
^{1/2}$. \ For any finite amplitude of oscillation, there will be non-harmonic
contributions from the electromagnetic forces on $e$ due to the charges $q$.
\ However, as the amplitude of oscillation becomes ever smaller, the
oscillation becomes ever-more-nearly harmonic. \ Furthermore, on
transformation to a new inertial frame $S^{\prime}$, the electromagnetic
system would have relativistic behavior in the limit of zero-oscillation
amplitude at a new Lorentz-transformed oscillation frequency $\omega
_{0}^{\prime}$. \ 

\subsubsection{Relativistic Harmonic Oscillator}

If we adopt this one-dimensional electromagnetic model for the oscillating
charge $e$ where the restoring force arises from the electrostatic field in
the rest frame of the charges $q$, then the relativistic equation for the
constrained motion for the charge $e$ in electromagnetic fields is
$d\mathbf{p}/dt=e\mathbf{E}\left(  \mathbf{r,}t\right)  .$ \ In the
approximation that the velocity of the charge $e$ relative to the charges $q$
is small compared to $c,$ one can approximate the relativistic change in
momentum as mass times acceleration, accurate through first order in the ratio
$v/c$. \ Thus we have \
\begin{equation}
\frac{d\mathbf{p}}{dt}=\frac{d}{dt}\left[  \frac{m\mathbf{v}}{\left(
1-v^{2}/c^{2}\right)  ^{1/2}}\right]  =\frac{d}{dt}\left[  m\mathbf{v}\left(
1+\frac{v^{2}}{2c^{2}}+...\right)  \right]  \approx\frac{d(m\mathbf{v})}{dt},
\end{equation}
provided that we can neglect terms of order $\left(  v/c\right)  ^{2}$. \ Our
electromagnetic oscillator can be treated as a relativistic system provided
that its oscillation velocity is small. \ The oscillator system is fully
relativistic only in the limit $v\rightarrow0$. \ 

\subsubsection{Radiation Equilibrium Beyond the Dipole Approximation}

In his textbook of classical electrodynamics, Jackson points out
that\cite{Jackson} \textquotedblleft Appreciable radiation in multiples of the
fundamental [oscillatory frequency] can occur because of relativistic effects
... even though the components of velocity are truly \textit{sinusoidal,} or
it can occur if the components of the velocity are not sinusoidal, even though
periodic.\textquotedblright\ \ All of the previous scattering calculations
have involved \textit{mechanical} motions which are \textquotedblleft not
sinusoidal, even though periodic,\textquotedblright\ as in our Eq.
(\ref{xtVV}). \ Thus the harmonics appear in the purely mechanical motion of
the charged particle. \ However, recent work by Huang and Batelaan\cite{HB}
regarding absorption of radiation at harmonics due to relativistic effects
suggested the possibility of determining radiation equilibrium not from
nonrelativistic mechanical motions but rather from purely relativistic effects
in classical electromagnetism. \ 

In problem 14 in Chapter 14, Jackson\cite{Jackson} asks a student to calculate
the radiation emitted at the harmonics $n\omega_{0}$ due to a purely
sinusoidal motion at $\omega_{0}$
\begin{equation}
x=D\cos\left[  \omega_{0}t+\phi\right]  .
\end{equation}
Thus the treatment of the multipole moments of the oscillating charge can be
extended beyond the dipole term to include the quadrupole moment of the
harmonic oscillator which will emit radiation at frequency $2\omega_{0}$, and
indeed can be extended to all higher multipoles. \ However, if the harmonic
oscillator can emit quadrupole radiation at the second harmonic $2\omega_{0}$
of the oscillator's natural frequency $\omega_{0}$, then it can also absorb
energy at the second harmonic. \ The force on the oscillator must be extended
beyond the dipole interaction $eE_{x}\left(  0,t\right)  $ to include the next
term in the expansion of the true force $eE_{x}\left(  x,t\right)  $ as
$eE_{x}\left(  0,t\right)  +ex\left[  \partial E_{x}(\widehat{i}x^{\prime
},t)/\partial x^{\prime}\right]  _{x^{\prime}=0}$. \ Thus a purely sinusoidal
mechanical motion of the oscillator combined with the \textit{relativistic}
radiation analysis can lead to a specific equilibrium spectrum of random
classical radiation. \ Here for the first time in the classical physics
\textit{scattering} literature, radiation equilibrium is determined not by
\textit{nonrelativistic} mechanical considerations but by
\textit{relativistic} electromagnetic aspects. \ 

\subsubsection{Zero-Point Radiation as the Oscillator-Scattering Equilibrium
Spectrum}

The calculation for the equilibrium radiation spectrum of a electromagnetic
charged harmonic oscillator of small amplitude has been carried
out.\cite{detailed} \ The equilibrium spectrum is a Lorentz-invariant
radiation spectrum. \ The Rayleigh-Jeans spectrum is not Lorentz invariant,
and so is \textit{not} the spectrum of radiation equilibrium for a charged
classical harmonic oscillator when treated beyond the dipole approximation for
the radiation interaction. \ Only Lorentz-invariant zero-point radiation will
serve as an equilibrium spectrum.

\subsubsection{Use of Nonrelativistic versus Relativistic Theory\ }

\ A harmonic oscillator is like a clock with frequency $\omega_{0}$. \ The
\textit{point} harmonic oscillator allows only a dipole connection to
radiation, and is in equilibrium with \textit{any} spectrum of isotropic
random radiation. \ In the dipole approximation, the radiation energy
$U_{rad}\left(  \omega_{0}\right)  $ merely determines the one number giving
the oscillator energy $U_{osc}\left(  \omega_{0}\right)  $. \ The point
oscillator clock itself can assume either Galilean or Lorentz transformation
properties when viewed in a new inertial frame. \ However, if the oscillator
motion has finite amplitude, then the analysis of the oscillator must choose
between the nonrelativistic and the relativistic transformations. \ If
\textit{nonrelativistic} mechanical motion beyond harmonic motion is involved,
then there is enormous flexibility in the choice of the mechanical interaction
potential $V\left(  x\right)  $; in equilibrium, the nonrelativistic nonlinear
oscillator, with a dipole connection to radiation, enforces a
frequency-independent constant energy on the radiation modes, corresponding to
the frequency-independent energy $k_{B}T$ allowed by the Wien law
(\ref{WienD}). \ On the other hand, if \textit{relativistic} electromagnetic
interactions are assigned to the harmonic oscillator, then the extension from
the dipole moment to the quadrupole moment and beyond allows no flexibility
whatsoever; the harmonic oscillator motion is completely determined, and the
equilibrium radiation spectrum assumes a Lorentz-invariant form, corresponding
to the zero-point energy limit $(1/2)\hbar\omega_{0}$ allowed by the Wien law
(\ref{WienD}).

\subsubsection{The Relativistic Limit and Classical Zero-Point Radiation}

The electromagnetic oscillator is fully relativistic only in the limit as its
velocity goes to zero. \ However, it turns out that the radiation balance at
each harmonic does not depend upon the actual oscillation amplitude for small
oscillations, provided that the amplitude is non-vanishing.\cite{detailed}
\ Thus the radiation equilibrium continues to hold without any change as the
amplitude of oscillation goes toward zero, and the oscillator motion becomes
ever closer to the relativistic limit. \ We expect that the radiation
equilibrium obtained from approximately relativistic oscillator motion will
hold even in the relativistic limit. \ 

If one uses the action-angle variables $w$ and $J$, then the Hamiltonian in
Eqs. (\ref{Ham}) and (\ref{HamJ}) makes no reference to the mass of the
oscillator or to the amplitude of oscillation, and the phase space
distribution for the oscillator given in Eq. (\ref{PHU}) becomes%
\begin{equation}
P_{osc}\left(  w,J\right)  =const\times\exp\left[  -\frac{J\omega_{0}}%
{U_{rad}\left(  \omega_{0}\right)  }\right]  . \label{PwJU}%
\end{equation}
This phase space distribution gives equilibrium for the oscillator at all the
harmonics $n\omega_{0}$ in a Lorentz-invariant radiation spectrum
$U_{rad}\left(  \omega\right)  $. \ If we take the Lorentz-invariant radiation
spectrum (required for full radiation equilibrium by the oscillator) as that
of classical electromagnetic zero-point radiation $U_{zpr\text{ zp}}\left(
\omega\right)  =\left(  1/2\right)  \hbar\omega$, then the oscillator phase
space (\ref{PwJU}) takes the form
\begin{equation}
P_{osc\text{ zp}}\left(  w,J\right)  =const\times\exp\left[  -\frac{J}%
{\hbar/2}\right]  , \label{PwJ}%
\end{equation}
where the oscillator frequency $\omega_{0}$ has cancelled out leaving a phase
space which independent of frequency. \ This phase space distribution for the
oscillator is the same as the phase space in Eq. (\ref{Pradzp}) for each mode
of electromagnetic radiation in the zero-point radiation field. \ The
equilibrium phase space distribution in Eq. (\ref{PwJ}) is invariant under an
adiabatic change of frequency $\omega_{0}$. \ 

\section{Scaling in Nonrelativistic Mechanics and in Relativistic
Electrodynamics}

In order to broaden our understanding of the contrasts arising in treatments
of classical radiation equilibrium, we now turn to the matter of scaling.
\ Scaling can be regarded as a change in the fundamental unit used to evaluate
some physical quantity, or as a multiplicative change in the physical quantity
itself. \ The contrasting scaling aspects of \textit{nonrelativistic}
classical mechanics and of \textit{relativistic} classical electrodynamics are
reflected in their determinations of thermal equilibrium. \ 

\subsection{Scaling Aspects of Nonrelativistic Mechanics}

\subsubsection{Separate Scalings in Length, Time, and Energy}

Because nonrelativistic mechanics involves no fundamental constants, it allows
separate scalings in length as $\sigma_{l}$, in time as $\sigma_{t}$, and in
energy as $\sigma_{U}$. \ For example, any nonrelativistic classical
mechanical system can be reimagined as a system where all the lengths are
twice as large ($\sigma_{l}=2)$, the times are three times as long
($\sigma_{t}=3)$, and the energies are four times as great ($\sigma_{U}=4$).
\ This flexibility arises since the only forms of energy are kinetic energy
(which can be scaled through the mass $m$), and potential energy $V(x,y,z)$
(which can be rescaled through the constants connecting distance to energy).
\ For a nonlinear mechanical oscillator of mass $m$, harmonic frequency
$\omega_{0}$, and nonlinear parameter $\alpha$, as in Eq. (\ref{Hnonlin}), the
quantities $m$, $\omega_{0}$, and $\alpha$ are all freely adjustable,
corresponding to allowing separate scalings $\sigma_{l}$, $\sigma_{t}$,
$\sigma_{U}$. \ 

\subsubsection{Thermal Equilibrium Reflects the Three Separate Scalings}

Equilibrium involving the interaction of classical mechanical systems must
reflect the three separate scalings allowed for nonrelativistic mechanical
systems. In an equilibrium situation for nonrelativistic physics, the energy
$k_{B}T$ must scale separately from the length and time parameters of the
mechanical systems.\ \ \textit{If we rescale the energy (by a factor }%
$\sigma_{U}$) \textit{of a mechanical system which is in equilibrium, then the
equilibrium of the system should not be disturbed by the energy rescaling.
}\ Furthermore, if we replace a member of a mechanical system by a new
mechanical member involving new lengths and times but with the same energy,
then the mechanical equilibrium will be unchanged because the total system
energy remains unchanged. \ If these scaling ideas are applied to the Wien-law
expression in Eq. (\ref{WienD}), they pick out only the equipartition limit
involving the energy-dependent temperature $T$ but having no dependence upon
the frequency-dependent $\omega$. \ Indeed the Boltzmann distribution reflects
this idea of a separate energy scaling. \ The Boltzmann probability on phase
space involves the mechanical energies of the constituent systems with no
other aspect involved. \ Interestingly enough, the Coulomb potential (which is
the only potential which can be extended to a fully \textit{relativistic}
classical electromagnetic theory) does not fit into the
\textit{nonrelativistic} Boltzmann analysis. \ 

The Rayleigh-Jeans spectrum reflects this nonrelativistic scaling pattern
where the energy $U_{rad}\left(  \omega,T\right)  =k_{B}T$ of each radiation
mode is the same, and is entirely independent of the frequency $\omega$ of the
radiation mode. \ It is interesting that in 1924, van Vleck\cite{VV359}
remarked with surprise regarding his nonrelativistic calculations that
\textquotedblleft...in a field of radiation whose specific energy [$\rho
_{rad}\left(  \omega,T\right)  =\left[  \omega^{2}/\left(  \pi^{2}%
c^{3}\right)  \right]  U_{rad}\left(  \omega,T\right)  $] does not vary with
the frequency, we have the rather surprising result that the mean absorption
is independent of the form of the force function $V\left(  x,y,z\right)  $
which holds the electron in the atom.\textquotedblright\ \ Apparently, the
idea of \textit{kinetic} energy equipartition, completely independent of the
force function $V\left(  x,y,z\right)  $, was a familiar and accepted idea of
nonrelativistic mechanics, whereas the directly-related result associated with
rates of energy absorption and loss from the radiation field treated in the
dipole approximation was regarded as \textquotedblleft
surprising.\textquotedblright\ \ Van Vleck's expressions for energy emission
and absorption in the Rayleigh-Jeans spectrum can be shown to give
\textit{kinetic} energy equipartition.\cite{oldqt}\ 

\subsubsection{No-Interaction Theorem in Relativistic Mechanics}

Many physicists\ are so accustomed to using nonrelativistic classical
mechanics with its freedom to choose interaction potentials $V\left(
x,y,z\right)  $ at will, that they are surprised that relativity places very
strong restrictions on systems. \ Indeed, most physicists have probably never
heard of the \textquotedblleft no-interaction theorem\textquotedblright\ of
Currie, Jordan, and Sudarshan\cite{CJS} which \textquotedblleft says that only
in the absence of direct particle interaction can Lorentz invariant systems be
described in terms of the usual position coordinates and corresponding
canonical momenta.\textquotedblright\cite{Goldstein}\ The standard
graduate-level mechanics text then simply dismisses relativistic ideas,
continuing, \textquotedblleft The scope of the relativistic Hamiltonian
framework is therefore quite limited and so for the most part we shall confine
ourselves to nonrelativistic mechanics.\textquotedblright\ \ \ 

\subsection{Scaling Aspects of Relativistic Electrodynamics}

\subsubsection{Single $\sigma_{ltU^{-1}}$-Scaling of Relativistic
Electrodynamics}

In contrast to the three separate scalings in length, time, and energy
appearing in nonrelativistic mechanics, electrodynamics allows only one single
scaling connecting together length, time, and energy. \ This situation arises
because classical electrodynamics involves several fundamental constants.
\ Length and time are coupled through the speed of light $c$, while energy and
length are coupled through the electronic charge $e,$ or through Stefan's
constant $a_{S}=\mathcal{U}/\left(  \mathcal{V}T^{4}\right)  $ related to the
total thermal-part $\mathcal{U}$\ of the radiation energy in a box of volume
$\mathcal{V}$. \ Therefore (relativistic) classical electrodynamics allows
only one scaling $\sigma_{ltU^{-1}}$ which preserves the values of these
fundamental constants.\cite{scale} \ We note that Planck's constant $\hbar$
has the same dimensions as $e^{2}/c$ and so is also unchanged under
$\sigma_{ltU^{-1}}$-scaling. \ Such $\sigma_{ltU^{-1}}$\textit{-scaling should
not disturb electromagnetic thermal equilibrium. \ }

In addition to preserving the fundamental constants $c,$ $e,$~and$~a_{S}$, the
$\sigma_{ltU^{-1}}$-scaling also preserves the form of Maxwell's equations.
\ \ Thus invariance under $\sigma_{ltU^{-1}}$-scaling holds for Gauss's law
$\nabla\cdot\mathbf{E}=4\pi\rho$ provided that the electric field $\mathbf{E}$
scales as charge divided by length squared. \ Faraday's law $\nabla
\times\mathbf{E=-}\left(  1/c\right)  \partial\mathbf{B/\partial}t$ satisfies
invariance under the scaling provided that the magnetic field $\mathbf{B}$
also scales as charge divided by length squared. \ Finally, the scaling for
the total energy $U$ in a volume $\mathcal{V}$, $U=u\mathcal{V}$, follows from
the scaling for the energy density $u=\left[  1/\left(  8\pi\right)  \right]
\left(  \mathbf{E}^{2}+\mathbf{B}^{2}\right)  $. \ 

\subsubsection{$\sigma_{ltU^{-1}}$-Scaling Allows a Function of $\omega/T$ for
Classical Thermal Radiation}

Lorentz-invariant classical zero-point radiation is $\sigma_{ltU^{-1}}%
$-invariant, and has no preferred length, time, or energy.\cite{zpscale}
\ Thus under a $\sigma_{ltU^{-1}}$-scale transformation, zero-point radiation
is mapped onto itself. \ For any radiation mode of frequency $\omega$, and
energy $U_{rad\text{ zp}}\left(  \omega\right)  =\left(  1/2\right)
\hbar\omega$, the $\sigma_{ltU^{-1}}$-scaling will carry the radiation mode
into a new radiation mode where the frequency is $\omega^{\prime}%
=\omega/\sigma_{ltU^{-1}}$ and the energy is $U^{\prime}=U^{\prime}%
/\sigma_{ltU^{-1}}$. \ But then we have $U_{rad\text{ zp}}^{\prime}\left(
\omega^{\prime}\right)  =U_{rad\text{ zp}}\left(  \omega\right)
/\sigma_{ltU^{-1}}=\left(  1/2\right)  \hbar\omega/\sigma_{ltU^{-1}}=\left(
1/2\right)  \hbar\omega^{\prime}=U_{rad\text{ zp}}\left(  \omega^{\prime
}\right)  $, so that the function connecting frequency and energy is
completely unchanged. \ Note that under this $\sigma_{ltU^{-1}}$-scale
transformation, the phase space distribution of each radiation mode remains
unchanged at $P_{rad\text{ zp}}(J_{rad})=const\times\exp\left[  -J_{rad}%
/\left(  \hbar/2\right)  \right]  $. \ On the other hand, thermal radiation at
temperature $T>0$ is not invariant under $\sigma_{ltU^{-1}}$-scaling. \ Since
temperature $T$ transforms as an energy, the information of Wien's law in Eq.
(\ref{WienD}) undergoes a $\sigma_{ltU^{-1}}$-scale transformation from
temperature $T$ to a new temperature $T^{\prime}=T/\sigma_{ltU^{-1}}$, while
making the ratio $\omega/T$ invariant, $\omega/T=\omega^{\prime}/T^{\prime}$.
\ Thus the $\sigma_{ltU^{-1}}$-scaling of relativistic classical
electrodynamics is consistent with Wien's law. \ 

\subsubsection{$\sigma_{ltU^{-1}}$-Scaling Allows Only Zero-Point Radiation
for the Relativistic Harmonic Oscillator}

Because of the existence of the fundamental constants $c$ and $e$ which
require the $\sigma_{ltU^{-1}}$-scaling of electromagnetism, a charged
(relativistic) electrodynamic system in thermal radiation can have at most one
freely-adjustable parameter for fixed temperature $T$. \ After the one
parameter of the charged system has been chosen, the conditions of thermal
equilibrium will determine the other physical quantities. \ For the
(relativistic) one-dimensional harmonic oscillator with its one scaling
parameter $\omega_{0}$, the oscillator energy $U_{osc}\left(  \omega
_{0}\right)  $ is determined by the initial conditions in connection with the
radiation spectrum at frequency $\omega_{0}$. \ 

In the calculation\cite{detailed} for the full electromagnetic radiation
equilibrium of an oscillator of small amplitude, purely electromagnetic
interactions are invoked to arrive at the same zero-point radiation spectrum
which is associated with adiabatic transformation of the oscillator.
\ However, the full Planck spectrum including zero-point radiation does
\textit{not} appear. \ Some physicists have taken this limitation as a sign
that there is something wrong with the analysis, and they feel justified in
their contentment with the erroneous claim that classical physics leads to the
Rayleigh-Jeans spectrum. \ However, the reason for the limitation in the
calculation is that the small-amplitude oscillator becomes a relativistic
system only at zero oscillation velocity, and therefore does not show the
variety of behavior of a fully relativistic electromagnetic system. \ Indeed,
the restriction of the (relativistic) oscillator to the
zero-oscillation-velocity limit necessarily excludes velocity-dependent
damping which is crucial for the low-frequency part of the Planck thermal
spectrum. \ 

\subsubsection{$\sigma_{ltU^{-1}}$-Scaling Allows a Function of $mc^{2}%
/\left(  k_{B}T\right)  $ for the Relativistic Classical Hydrogen Atom}

\paragraph{Scaling for the Classical Hydrogen Atom}

The limiting considerations for the (approximately relativistic) harmonic
oscillator do not appear for the relativistic classical hydrogen atom. \ The
classical hydrogen atom consists of a point charge $e$ of mass $m$ in a
Coulomb potential $V\left(  r\right)  =-e^{2}/r$ in the presence of random
classical radiation. \ This system is fully relativistic when considered as
part of classical electrodynamics. The one freely-adjustable scaling parameter
is the mass $m.$ Using this mass and the fundamental constants $e$ and $c$,
one can form the energy $mc^{2}$, the length $e^{2}/\left(  mc^{2}\right)  $,
and time $e^{2}/\left(  mc^{3}\right)  $. \ Under a $\sigma_{ltU^{-1}}$-scale
transformation, the mass $m$ is carried into mass $m^{\prime}=m/\sigma
_{ltU^{-1}}$, but velocities are unchanged. \ Since classical zero-point
radiation is $\sigma_{ltU^{-1}}$-scale invariant, then the classical hydrogen
atom in zero-point radiation will have its equilibrium phase space unchanged
by the transformation. \ Since velocities are unchanged under the
transformation, this fits with the absence of velocity-dependent damping in
zero-point radiation; only acceleration-dependent radiation damping is
involved in equilibrium. \ It is familiar that for circular orbits, the speed
of a particle in a circular orbit is $v=e^{2}/J$ where $J$ is the angular
momentum. \ Under a $\sigma_{ltU^{-1}}$-scale transformation, the quantities
$v,~e,$ and $J$ all remain unchanged. \ 

\paragraph{Scaling for Hydrogen in Zero-Point Radiation}

In zero-point radiation, the scale of the random radiation is given by $\hbar$
which has the same dimensions as $e^{2}/c$. \ Thus in the limit which removes
all factors of $c$, we find that the classical hydrogen atom in zero-point
radiation has a typical length $\left[  e^{2}/(mc^{2})\right]  \left[  \hbar
c/e^{2}\right]  ^{2}=\hbar^{2}/\left(  me^{2}\right)  $, a typical time
$\left[  e^{2}/(mc^{3})\right]  \left[  \hbar c/e^{2}\right]  ^{3}=\hbar
^{3}/\left(  me^{4}\right)  $, and a typical energy $mc^{2}\left[
e^{2}/\left(  \hbar c\right)  \right]  ^{2}=me^{4}/\hbar^{2}$. \ These
quantities, which involve no factors of $c$, are familiar from the Bohr model
of hydrogen. \ They also appear in work involving both numerical simulations
and the Fokker-Planck equation for the classical hydrogen atom in classical
zero-point radiation.\cite{CZ}

\paragraph{Scaling for Hydrogen in Thermal Radiation with $T>0$}

If the classical hydrogen atom is in equilibrium with thermal radiation at
temperature $T>0$, then there will be velocity-dependent damping and a
$\sigma_{ltU^{-1}}$-scale transformation will indeed change the system. \ The
relativistic particle in a Coulomb potential allows a phase space distribution
dependent upon both the mass $m$, and also the fundamental constants $e$ and
$c$. \ Thus the phase space for the Coulomb situation can involve the
characteristic energy ratio $mc^{2}/\left(  k_{B}T\right)  $ which is freely
adjustable in both $m$ and in $T.$\ This freely-adjustable ratio allows the
possibility of a transition between a high-frequency region dominated by
acceleration-dependent radiation damping and a low-frequency region dominated
by velocity-dependent damping. \ 

For the Coulomb potential, when the orbital radius is small, the velocity is
high and the frequency is high, corresponding to the zero-point radiation part
of the spectrum where acceleration-dependent radiation damping dominates and
velocity-dependent damping is minor. On the other hand, when the orbital
radius is large, the velocity and frequency are low, and acceleration-based
radiation damping is also low. \ In this low-frequency region,
velocity-dependent damping due to thermal radiation for $T>0$ might well
dominate. \ This situation can be illustrated\cite{unfam} by considering
\ relativistic \textit{circular} orbits characterized by angular momentum $J$
where the velocity is $v=e^{2}/J,$ the energy is $U(J)=mc^{2}\sqrt{1-\left[
e^{2}/(Jc)\right]  ^{2}}$, the radius is $r=\left[  e^{2}/\left(
mc^{2}\right)  \right]  \left(  Jc/e^{2}\right)  ^{2}\sqrt{1-\left[
e^{2}/(Jc)\right]  ^{2}}$ , and the frequency is $\omega=\partial U/\partial
J=\left(  mc^{3}/e^{2}\right)  \left[  e^{2}/\left(  Jc\right)  \right]
^{3}/\sqrt{1-\left[  e^{2}/(Jc)\right]  ^{2}}$. \ The ratio of orbital
frequency $\omega$ to temperature $T$ can be written as
\begin{equation}
\frac{\hbar\omega}{k_{B}T}=\frac{mc^{2}}{k_{B}T}\left(  \frac{\hbar c}{e^{2}%
}\right)  \left(  1-\left(  \frac{e^{2}}{Jc}\right)  ^{2}\right)
^{-1/2}\left(  \frac{e^{2}}{Jc}\right)  ^{3}.\label{ratio}%
\end{equation}
We see from Eq. (\ref{ratio}), that for changing temperature $T$ and for fixed
mass $m$ and fixed angular momentum $J$ (which corresponds to fixed frequency
$\omega$), the system changes from the high-frequency region of the thermal
spectrum $\hbar\omega/k_{B}T>>1,$ to the low-frequency region $\hbar
\omega/k_{B}T<<1$ exactly in unison with the ratio $mc^{2}/\left(
k_{B}T\right)  .$ \ The transition between the two regions of behavior is
mediated by the ratio $mc^{2}/\left(  k_{B}T\right)  $. \ An analogous ratio
does not exist for the (approximately relativistic) harmonic oscillator. \ 

\section{Miscellaneous Aspects of Thermal Equilibrium}

\subsection{Full Thermal Equilibrium Requires Both Velocity-Dependent Damping
and Acceleration-Dependent Damping}

Indeed it is interesting to see that within classical physics, thermal
radiation equilibrium for non-zero temperature $T>0$ requires both
acceleration-dependent radiation damping and velocity-dependent damping. \ The
Einstein-Hopf analysis\cite{EH}\cite{Milonni} of 1910 considered a particle of
large mass $M$ containing an electric dipole oscillator which experienced
velocity-dependent damping when moving through random radiation, but there
were no non-radiation forces and so the particle motion experienced no
acceleration-dependent damping. \ Assuming that the average kinetic energy of
the one-dimensional motion of the large mass $M$ was $\left(  1/2\right)
k_{B}T$ , the Einstein-Hopf analysis led to the low-frequency,
velocity-dependent, Rayleigh-Jeans part of the Planck law for thermal
radiation equilibrium. \ Indeed, the velocity-dependent damping without
acceleration-dependent radiation damping is analogous to the situation which
arises in \textit{nonrelativistic} mechanics where thermal equilibrium depends
entirely upon velocity-dependent damping, and no acceleration-dependent
damping can appear. \ It was only later when the Einstein-Hopf analysis was
extended\cite{without} to also include acceleration-dependent radiation
damping of the particle motion (arising from non-radiation forces on the
particle) that the classical Einstein-Hopf analysis led to the full Planck
spectrum including zero-point radiation. \ Furthermore, the inclusion of
classical zero-point radiation in the Planck spectrum suggests the possibility
of superfluid-like behavior for an Einstein-Hopf particle at low
temperatures.\cite{super} \ 

\subsection{Information Contained within the Rayleigh-Jeans and Zero-Point
Radiation Spectra}

\subsubsection{Zero-Point Spectrum Has Least Information}

Zero-point radiation involves the spectrum of random radiation with least
possible information. \ The spectrum is Lorentz invariant and has no preferred
inertial frame, and hence no velocity-dependent damping. \ The spectrum is
also scale invariant, and has no preferred length or time or energy. \ Even in
curved spacetime, zero-point radiation has correlation functions which involve
only the geodesic separations between the spacetime points where the
correlation is evaluated. \ 

\subsubsection{Information in the Rayleigh-Jeans Spectrum}

The Rayleigh-Jeans spectrum contains more information than the classical
zero-point radiation spectrum. \ Both spectra are characterized by a single
parameter: $\hbar$ for zero-point radiation and $T$ for the Rayleigh-Jeans
spectrum. \ For the Rayleigh-Jeans spectrum, the temperature $T$ is
freely-adjustable, and the energy per normal mode is $U_{rad~\text{RJ}}\left(
\omega,T\right)  =k_{B}T$ \ at any frequency $\omega.$ \ However, the
Rayleigh-Jeans spectrum determines exactly one preferred inertial frame in
which it is isotropic. \ Furthermore, at fixed temperature $T$, one can
determine whether one frequency is larger than another frequency by comparing
their phase space distributions. \ Thus the phase space distribution for
electromagnetic radiation in the Rayleigh-Jeans spectrum is $P_{rad\text{ RJ}%
}\left(  J,\omega,T\right)  =const\times\exp\left[  -J\omega/\left(
k_{B}T\right)  \right]  $ so that a larger value of frequency $\omega$
corresponds to a phase space distribution more concentrated near $J=0$. \ On
the other hand, the classical zero-point radiation spectrum $U_{rad\text{ zp}%
}\left(  \omega\right)  =\left(  1/2\right)  \hbar\omega$ has a phase space
distribution which is the same for radiation modes of any frequency $\omega$,
$P_{rad\text{ zp}}\left(  J,\omega\right)  =const\times\exp\left[  -J/\left(
\hbar/2\right)  \right]  .$ \ Zero-point radiation has no preferred inertial
frame and is isotropic in every inertial frame. \ If a harmonic oscillator of
small amplitude were to scatter zero-point radiation so as to enforce any
radiation spectrum other than the zero-point radiation spectrum, the
oscillator would impose a preferred inertial frame upon the scattered
radiation. \ 

\subsubsection{Thermal Radiation Spectrum as Giving Least Information in the
Presence of Zero-Point Radiation}

\textit{Thermal} radiation at $T>0$ involves a finite amount of energy spread
over the modes of the zero-point radiation spectrum and so introduces a
preferred inertial frame. \ The spectrum of zero-point radiation acts as a
noise spectrum into which the finite amount of thermal energy is introduced,
and the spectrum of thermal radiation represents the minimal information
consistent with the finite total available thermal energy and the divergent
Lorentz-invariant spectrum of zero-point radiation. \ The entropy of thermal
radiation at frequency $\omega$ is determined by comparing the amount of
thermal radiation to the amount of zero-point radiation at that
frequency.\cite{BB2018}\ 

\subsection{Inconsistent Mixtures of Nonrelativistic and Relativistic Physics}

The derivations of the Rayleigh-Jeans law for thermal radiation given in the
textbook literature arise from attempts to combine \textit{nonrelativistic}
mechanics with \textit{relativistic} electrodynamics. \ Such inconsistent
mixtures of nonrelativistic and relativistic theories satisfy neither Galilean
invariance nor Lorentz invariance. Indeed, the results of the analysis depend
upon the inertial frame in which the analysis is done. \ For particles
undergoing point collisions, this situation has already been exhibited
explicitly in the literature.\cite{mixed} \ Thus if one considers the point
collision between two particles, using relativistic physics for one and
nonrelativistic physics for the other, one can solve for the motion of the
particles after the collision by using the energy and momentum conservation
laws in any inertial frame; however, the results satisfy neither the
\textit{nonrelativistic} conservation law for the uniform motion of the center
of mass nor the \textit{relativistic} conservation law for the uniform motion
of the center of energy, and the results indeed depend upon the inertial frame
chosen for the analysis. \ 

\section{Closing Summary}

\subsection{Two Natural Spectra Associated with Harmonic Oscillators}

Associated with harmonic oscillators, there are two spectra of random
radiation which arise in a natural manner. \ One of these is the
Rayleigh-Jeans spectrum which is associated with the equipartition ideas of
nonrelativistic classical mechanics and is based upon velocity-dependent
damping. \ The other is the classical zero-point radiation spectrum which
arises from an adiabatic transformation of oscillator frequency and is
associated with relativistic classical electrodynamics which includes
acceleration-dependent damping. \ 

\subsection{Oscillators in Translational Motion and Two Forms of Damping}

If one considers not a stationary oscillator but rather an oscillator which is
free to move in over-all translation, then one can obtain either the
Rayleigh-Jeans spectrum (obtained for a situation with velocity-dependent
damping on the oscillator but no acceleration-dependent damping by Einstein
and Hopf\cite{EH}) or the full Planck spectrum with zero-point radiation (if
one allows both types of damping on the oscillator\cite{without}). \ 

\subsection{Thermal Radiation Equilibrium from Scattering}

A charged harmonic oscillator at rest with its electromagnetic interactions
limited to the dipole approximation is in equilibrium with \textit{any}
spectrum of radiation; the energy of the oscillator merely matches the energy
of the radiation modes at the natural frequency of the oscillator. \ If we
wish to go beyond this ambiguous radiation situation and to determine a
preferred equilibrium radiation spectrum under \textit{scattering}, then there
are two natural \textit{scattering} extensions. \ If we continue the
limitation on the electromagnetic interactions to the dipole approximation but
go to a more complex \textit{nonrelativistic} mechanical motion, then we
arrive at the Rayleigh-Jeans spectrum. \ On the other hand, if we treat the
charged harmonic oscillator as a fully electromagnetic system but with only
\textit{approximately-relativistic} harmonic oscillator mechanical behavior,
then we arrive at classical zero-point radiation as the preferred radiation
spectrum. \ Scattering by a \textit{fully relativistic} electromagnetic
system, such as the classical hydrogen atom, has the possibility of giving the
full Planck radiation spectrum with zero-point radiation as the equilibrium
classical radiation spectrum. \ 

\subsection{Relativity and Thermal Radiation}

If one reads James Jeans' \textit{Report on Radiation and the Quantum-Theory}
published in 1914 or if one reads the modern physics textbooks published in
the last few years, one would find no inkling that special relativity might
have some relevance to the blackbody radiation problem. \ However, the
equilibrium radiation spectrum is determined by assumptions on both the
mechanical motion (whether nonrelativistic or relativistic) and the
relativistic electromagnetic interactions. \ In any case, the radiation
equilibrium of the fully electromagnetic charged harmonic oscillator of small
amplitude does \textit{not} lead to the Rayleigh-Jeans spectrum. \ It is an
outright error to claim that classical physics leads inevitably to the
Rayleigh-Jeans spectrum.


\begin{thebibliography}{99}                                                                                               %


\bibitem {modern}See for example, R. Eisberg and R. Resnick, \textit{Quantum
Physics of Atoms, Molecules, Solids, Nuclei, and Particles }2nd ed. (Wiley,
New York 1985); K. S. Krane, \textit{Modern Physics}, 2nd ed. (Wiley, New York
1996); R. Taylor, C. D. Zafiratos, and M. A. Dubson, \textit{Modern Physics
for Scientists and Engineers}, 2nd ed. (Pearson, New York, 2003); S. T.
Thornton and A. Rex, \textit{Modern Physics for Scientists and Engineers}, 4th
ed. (Brooks/Cole, Cengage Learning, Boston, MA, 2013).

\bibitem {Jeans}J. H. Jeans, \textit{Report on Radiation and the Quantum
Theory} (www.ForgottenBooks.org, 2013). \ This is a reproduction from the
report originally published in 1914.

\bibitem {BB2018}T. H. Boyer, \textquotedblleft Blackbody radiation in
classical physics: A historical perspective,\textquotedblright\ Am. J. Phys.
\textbf{86}, 495-509 (2018). \ 

\bibitem {CJS}D. G. Currie, T. F. Jordan, and E. C. G. Sudarshan,
\textquotedblleft Relativistic Invariance and Hamiltonian theories of
interacting particles,\textquotedblright\ Rev. Mod. Phys. \textbf{35}, 350-375 (1963).

\bibitem {Wein}T. H. Boyer, \textquotedblleft Thermodynamics of the harmonic
oscillator: Wien's displacement law and the Planck spectrum,\textquotedblright%
\ Am. J. Phys. \textbf{71}, 866-870 (2003).

\bibitem {third}T. H. Boyer, \textquotedblleft Thermodynamics of the harmonic
oscillator: derivation of the Planck blackbody spectrum from pure
thermodynamics,\textquotedblright\ Eur. J. Phys. \textbf{40,} 025101(16pp)
(2019). \ 

\bibitem {Planck}See for example, M. Planck, \textit{The Theory of Heat
Radiation} (Dover, New York 1959).

\bibitem {M63}T. W. Marshall, \textquotedblleft Random
electrodynamics,\textquotedblright\ Proc. R. Soc. \textbf{A276}, 475-491 (1963).

\bibitem {Rev75}T. H. Boyer, \textquotedblleft Random electrodynamics: The
theory of classical electrodynamics with classical electromagnetic zero-point
radiation,\textquotedblright\ Phys. Rev. D \textbf{11}, 790-808 (1975).

\bibitem {Lav}B. H. Lavenda, \textit{Statistical Physics: A Probabilistic
Approach} (Wiley, New York 1991), pp. 73-74.

\bibitem {ZPE}T. H. Boyer, \textquotedblleft Understanding zero-point energy
in the context of classical electromagnetism,\textquotedblright\ Eur. J. Phys.
\textbf{37}, 055206(14) (2016). \ 

\bibitem {Rev}A review of the work on classical electromagnetic zero-point
radiation up to 1996 is provided by L. de la Pena and A. M. Cetto, \textit{The
Quantum Dice - An Introduction to Stochastic Electrodynamics} (Kluwer
Academic, Dordrecht 1996). \ For a more recent short review, see T. H. Boyer,
\textquotedblleft Stochastic Electrodynamics: The Closest Classical
Approximation to Quantum Theory,\textquotedblright\ Atoms \textbf{7}(1), 29-39 (2019).

\bibitem {const}T. H. Boyer, \textquotedblleft The contrasting roles of
Planck's constant in classical and quantum theories,\textquotedblright\ Am. J.
Phys. \textbf{86}, 280-283 (2018).

\bibitem {EH}A. Einstein and L. Hopf, \textquotedblleft Statistische
Untersuchung der Bewegung eines Resonators in einem
Strahlungsfeld,\textquotedblright\ Annalen der Physik (Leipzig) \textbf{33},
1105-1115 (1910).

\bibitem {Milonni}For modern notation, see P. W. Milonni, \textit{The Quantum
Vacuum: An Introduction to Quantum Electrodynamics} (Academic Press,
Boston1994), pp. 11-14.

\bibitem {VV}J. H. van Vleck, \textquotedblleft The absorption of radiation by
multiply periodic orbits, and its relation to the correspondence principle and
the Rayleigh-Jeans law: Part II. Calculation of absorption by multiply
periodic orbits,\textquotedblright\ Phys. Rev. \textbf{24}, 347-365 (1924);
\textquotedblleft A correspondence principle for absorption,\textquotedblright%
\ Jour. Opt. Soc. Amer. \textbf{9}, 27-30 (1924). \ 

\bibitem {nonlin}T. H. Boyer, \textquotedblleft Equilibrium of random
classical electromagnetic radiation in the presence of a nonrelativistic
nonlinear electric dipole oscillator,\textquotedblright\ Phys. Rev. D
\textbf{13}, 2832-2845 (1976).

\bibitem {Blanco}R. Blanco, L. Pesquera, and E. Santos, \textquotedblleft
Equilibrium between radiation and matter for classical relativistic
multiperiodic systems. \ Derivation of Maxwell-Boltzmann distribution from
Rayleigh-Jeans spectrum,\textquotedblright\ Phys. Rev. D\textbf{27}, 1254-1287
(1983); \textquotedblleft Equilibrium between radiation and matter for
classical relativistic multiperiodic systems. II. Study of radiative
equilibrium with Rayleigh-Jeans radiation,\textquotedblright\ Phys. Rev. D
\textbf{29}, 2240-2254 (1984).

\bibitem {Born}M. Born, \textit{The Mechanics of the Atom} (Ungar, New York
1970), pp. 66-71.

\bibitem {Jackson}J. D. Jackson, \textit{Classical Electrodynamics }3rd ed.
(John Wiley \& Sons, New York, 1999), p. 704, problem 14.22.

\bibitem {HB}W. C-W. Huang and H. Batelaan, \textquotedblleft Discrete
Excitation Spectrum of a Classical Harmonic Oscillator in Zero-Point
Radiation,\textquotedblright\ Found. Phys. \textbf{45}, 333-353 (2015).

\bibitem {detailed}T. H. Boyer, \textquotedblleft Equilibrium for classical
zero-point radiation: detailed balance under scattering by a classical charged
harmonic oscillator,\textquotedblright\ J. Phys. Commun. \textbf{2},
105014(17) (2018).

\bibitem {VV359}See ref. 16, p. 359. \ Van Vleck's comment involves radiation
absorption but is related to nonrelativistic kinetic energy and to the
nonrelativistic Larmor formula for radiation emission. \ Thus for a general
oscillation such as suggested by our Eq. (\ref{xtVV}), the average kinetic
energy in the $n^{th}$ harmonic involves $\left(  n\omega_{1}\right)
^{2}D_{n}^{2}$ whereas the average radiation emission by the $n^{th}$ harmonic
is $\left(  n\omega_{1}\right)  ^{2}$ times as large, involving $\left(
n\omega_{1}\right)  ^{4}D_{n}^{2}=$ $\left(  n\omega_{1}\right)  ^{2}\left[
\left(  n\omega_{1}\right)  ^{2}D_{n}^{2}\right]  $.

\bibitem {oldqt}T. H. Boyer, \textquotedblleft Statistical equilibrium of
nonrelativistic multiply periodic classical systems and random classical
electromagnetic radiation,\textquotedblright\ Phys. Rev. A \textbf{18},
1228-1237 (1978).

\bibitem {scale}T. H. Boyer, \textquotedblleft Scaling symmetries of
scatterers of classical zero-point radiation,\textquotedblright\ J. Phys. A:
Math. Theor. \textbf{40}, 9635-9642 (2007); \ \textquotedblleft Scaling
symmetry and thermodynamic equilibrium for classical electromagnetic
radiation,\textquotedblright\ Found. Phys. \textbf{19}, 1371-1383 (1989).

\bibitem {Goldstein}H. Goldstein, C. Poole, and J. Safko, \textit{Classical
Mechanics }3rd ed. (Addison-Wesley, New York, 2002), p. 353.

\bibitem {zpscale}T. H. Boyer, \textquotedblleft Conformal Symmetry of
Classical Electromagnetic Zero-Point Radiation,\textquotedblright\ Found.
Phys. \textbf{19}, 349-365 (1989).

\bibitem {CZ}For work on the classical hydrogen atom in classical zero-point
radiation, see D. C. Cole and Y. Zou, \textquotedblleft Quantum Mechanical
Ground State of Hydrogen Obtained from Classical
Electrodynamics,\textquotedblright\ Phys. Lett. A \textbf{317}, 14-20 (2003),
and T. H. Boyer, \textquotedblleft Relativity and Radiation Balance for the
Classical Hydrogen Atom in Classical Electromagnetic Zero-Point
Radiation,\textquotedblright\ Eur. J. Phys. \textbf{42}, 025205 (24pp) (2021).

\bibitem {unfam}T. H. Boyer, \textquotedblleft Unfamiliar trajectories for a
relativistic particle in a Kepler or Coulomb potential,\textquotedblright\ Am.
J. Phys. \textbf{75}, 992-997 (2004).

\bibitem {without}T. H. Boyer, \textquotedblleft Derivation of the Blackbody
Radiation Spectrum without Quantum Assumptions,\textquotedblright\ Phys. Rev.
\textbf{182}, 1374-1383 (1969).

\bibitem {super}T. H. Boyer, \textquotedblleft Particle Brownian motion due to
random classical radiation: Superfluid-like behavior in classical zero-point
radiation,\textquotedblright\ Eur. J. Phys. \textbf{41}, 055103 (17pp) (2020).

\bibitem {mixed}T. H. Boyer, \textquotedblleft Illustrating some implications
of the conservation laws in relativistic mechanics,\textquotedblright\ Am. J.
Phys. \textbf{77}, 562-569 (2009).
\end{thebibliography}
\end{document}